\documentclass[11pt,twoside]{article}
\usepackage{asp2014}
\usepackage{aas_macros}
\aspSuppressVolSlug
\resetcounters

\begin{document}

\title{WTF? Discovering the Unexpected in next-generation radio continuum surveys}
\author{Evan~Crawford$^1$, Ray~P.~Norris$^{1,2}$, and Kai Polsterer$^3$
\affil{$^1$Western Sydney University, Penrith, NSW, Australia; \email{e.crawford@westernsydney.edu.au}}
\affil{$^2$CSIRO Astronomy \& Space Science, Epping, NSW, Australia \email{raypnorris@gmail.com}}
\affil{$^3$Heidelberg Institute for Theoretical Studies, Heidelberg, Germany \email{Kai.Polsterer@h-its.org}}
}

\paperauthor{Evan Crawford}{e.crawford@westernsydney.edu.au}{}{Western Sydney University}{School of Computing, Engineering and Mathematics}{Penrith}{NSW}{2751}{Australia}
\paperauthor{Ray Norris}{raypnorris@gmail.com}{0000-0002-4597-1906}{CSIRO}{Astronomy\&SpaceScience}{Epping}{NSW}{2121}{Australia}
\paperauthor{Kai Polsterer}{Kai.Polsterer@h-its.org}{}{Heidelberg Institute for Theoretical Studies}{}{Heidlberg}{}{69118}{Germany}

\begin{abstract}
Most major discoveries in astronomy have come from unplanned discoveries made by surveying the Universe in a new way, rather than by testing a hypothesis or conducting an investigation with planned outcomes. Next generation radio continuum surveys such as the Evolutionary Map of the Universe (EMU: the radio continuum survey on the new Australian SKA Pathfinder telescope), will significantly expand the volume of observational phase space, so we can be reasonably confident that we will stumble across unexpected new phenomena or new types of object. However, the complexity of the instrument and the large data volumes mean that it may be non-trivial to identify them. On the other hand, if we don't, then we may be missing out on the most exciting science results from EMU. We have therefore started a project called ``WTF'', which explicitly aims to mine EMU data to discover unexpected science that is not part of our primary science goals, using a variety of machine-learning techniques and algorithms. Although targeted specifically at EMU, we expect this approach will have broad applicability to astronomical survey data.
\end{abstract}

\section{Introduction}
Major scientific discoveries,  such as the discovery of the  Higgs Boson \citep{Aad2012}, are often made by testing a hypothesis or conducting an investigation with planned outcomes. However, most major discoveries in astronomy are unplanned and unexpected \citep{Harwit1981, 2004NewAR..48.1551W, 2009arad.workE...7E, 2015aska.confE..86N, Norris2015b}. Typically these occur as the result of building larger telescopes, or  opening up a new window of the electromagnetic spectrum. More generally, we may define a phase space whose axes correspond to observable quantities. Some parts of this phase space (e.g. bottom right of Fig.~1) have been well-observed and have already yielded their discoveries, whereas some parts of this space (e.g. top left of Fig.~1) have not yet been observed, and may contain new discoveries that are available to new instruments sampling that region of  phase space. Virtually all ``accidental'' or ``serendipitous'' discoveries result  from observing a new part of this phase space \citep{Norris2015b}. The case study of the Nobel-prize-winning discovery of pulsars by Jocelyn Bell is instructive. A talented and persistent PhD student studying interstellar scintillation (and thus expanding the observational phase space), and who knew her instrument intimately, recognised that `bits of scruff' on the chart recorder could not be terrestrial interference, but represented a new type of astronomical object \citep{2009arad.workE..14B}. As a result, she discovered pulsars.

\begin{figure}
  \begin{minipage}[c]{8.25cm}
    \includegraphics[width=8.25cm]{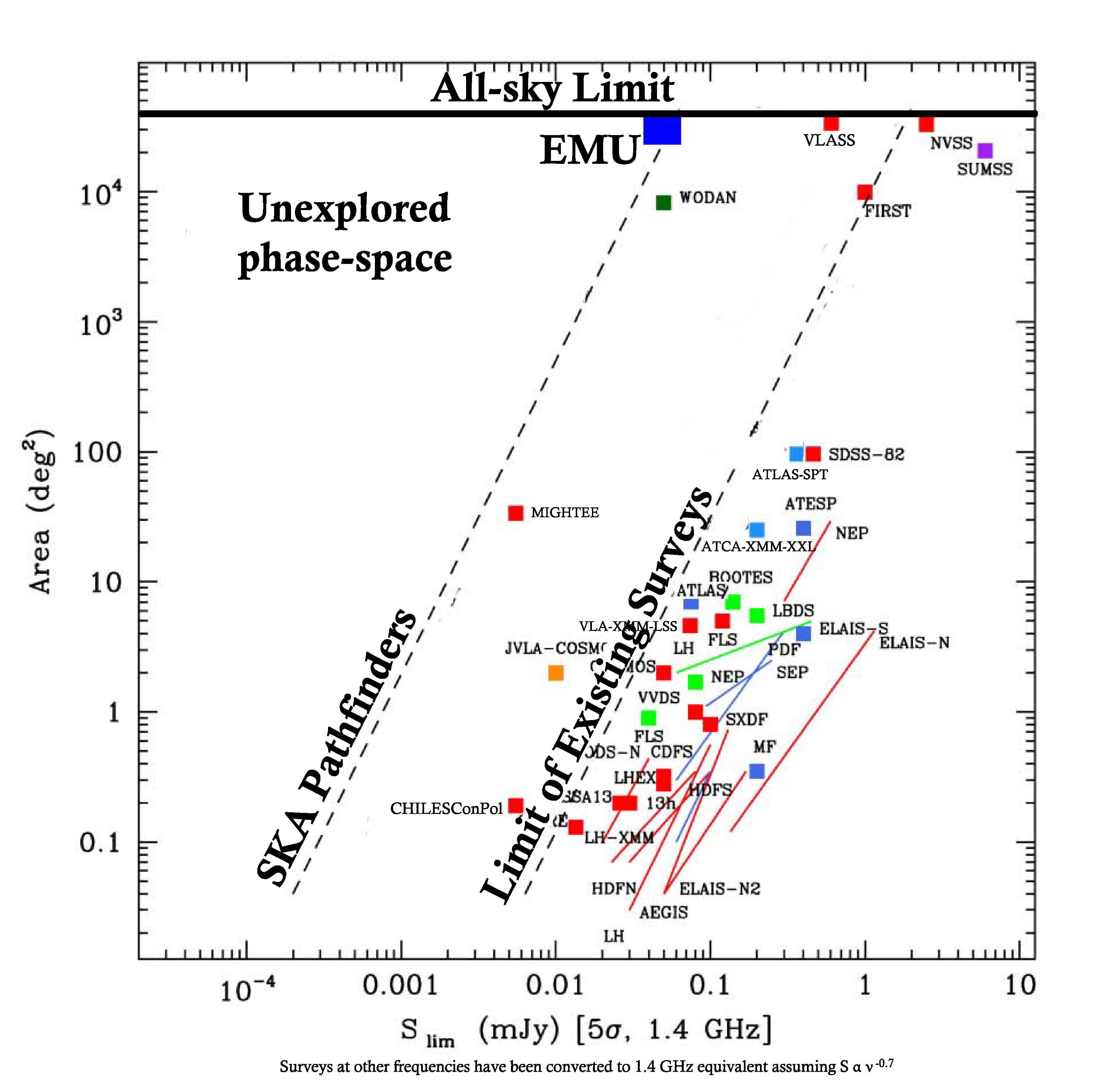}
  \end{minipage}\hfill
  \begin{minipage}[c]{5.75cm}
    \caption{
       {\scriptsize Comparison of  existing and planned deep 20 cm radio continuum surveys. EMU (and the smaller WODAN and MIGHTEE surveys) are the only surveys that extend an order of magnitude to the left of existing surveys, exploring new observational phase space. Surveys at other wavelengths (e.g. MWA, \mbox{LOFAR}) are not shown, but will detect significantly fewer galaxies than EMU. The horizontal axis shows the  5-$\sigma$ sensitivity, and the vertical axis shows the sky coverage. The right-hand diagonal dashed line shows the approximate envelope of existing surveys, which is largely determined by the availability of telescope time.}
}
\end{minipage}
\end{figure}

The Evolutionary Map of the Universe \citep[EMU][]{2011PASA...28..215N}, the radio continuum survey for the new Australian SKA Pathfinder
telescope \citep[ASKAP][]{2008ExA....22..151J}, will produce a deep ($\sim\ 10 \mu\mathrm{Jy/beam}$) radio-continuum map of the sky. EMU is expected to detect around 70 million galaxies, compared to the total of about 2.5 million so far discovered over the entire history of radio-astronomy.
 
 Figure 1 shows that EMU will significantly expand the volume of observational phase space, so we can be reasonably confident that it will stumble across unexpected new phenomena or new types of object. Although it is impossible to predict their nature, we might reasonably expect new classes of galaxy, or large-scale phenomena affecting galaxies in some parts of the sky. However, the complexity of the instrument and the large data volumes mean that it may be non-trivial to identify them. For example, a present-day Jocelyn Bell is unlikely to understand the instrument well enough to distinguish astrophysical phenomena from instrumental effects, and would not be able to sift through the petabytes by hand, searching for something unusual. On the other hand, failure to identify unexpected effects may mean missing out on the most important science to emerge from ASKAP. It is therefore necessary to plan explicitly to build techniques to make unexpected discoveries, rather than hoping to stumble across them.  

We have therefore started a project called ``WTF'', which explicitly aims to mine EMU data to discover unexpected science that is not part of our primary science goals, using a variety of machine-learning techniques and algorithms. Although targeted specifically at EMU, we expect this approach to have broad applicability to astronomical survey data.



\section{Widefield ouTlier Finder}
\textbf{W}idefield ou\textbf{T}lier \textbf{F}inder (WTF) is a pilot project to test and implement techniques for mining large data sets for unexpected discoveries. 
WTF takes the overall approach initially of constructing large data sets containing simulated discoveries (named ``eggs'') and then inviting competing algorithms and techniques to find these eggs in a series of `data challenges'. 
 Visualisation tools will also be developed to aid understanding of the process and its results.
 The data include both real and simulated data, and include both images (which may include spectral shape and polarisation data) and tabular data, and will include the multiwavelength identifications and other properties where available.
 
Different algorithms are then applied in blind tests to see which are most successful at finding eggs. This process includes inviting external collaborators to try their algorithms on the data. Because some of the data is real data that has not been mined, there is a chance, even at this early stage, that the teams may make genuine unexpected discoveries. In later stages of the project, the most successful techniques will be used to search for discoveries in real EMU data. The large data volumes necessitate  automatic processing, with minimal human interaction, providing a challenge well-suited to machine learning techniques.

We are implementing WTF initially as a collaborative environment on the 
Amazon Web Services\footnote{\url{http://aws.amazon.com}} platform, which provides disk space, processing capacity, and infrastructure. The AWS platform is well-suited to this project, in providing a collaborative research environment, with disk and processing capacity that can be varied  to suit the demand.

We expect many different algorithms to be applied to the task of finding the eggs. For example, one approach may look for groups of properties in an n-dimensional plot with axes such as flux density, spectral index, and IR-to-radio ratio.  Some groups will correspond to known types of object (e.g.\ stars, galaxies, quasars) but unexpected groups may correspond to unknown classes.
More sophisticated approaches will use machine-learning algorithms such as neural nets, dimensionality reduction (self-organ\-ising maps, autoencoding, isoplanar mapping, TSNE, etc.), clustering (k-means, db-scan, Birch), outlier detection, and Bayesian approaches.  Although targeted  at EMU, such approaches will be widely applicable to astronomical survey data.

It is likely that different algorithms will be most suited for particular challenges, and so, even in the latest stages of this project, when EMU data is mined for unexpected discoveries, a variety of different algorithms will still be used. Furthermore, it is likely that new algorithms and new approaches will evolve during the lifetime of this project, so a cyclic process is envisaged where the data challenges are followed by periods of algorithm development followed by further challenges.

We intend to run the WTF project in 4 phases, as follows. Only the first phase is currently funded.

\begin{itemize}
\item Phase 1: Set up infrastructure, and initial challenges consisting of data (images or tables) with embedded ``eggs''; apply the first challenger algorithms to debug and refine the infrastructure. Validate the architecture using in-house challengers; test that the infrastructure is extendable to a larger number of users.
\item Phase 2: Test different approaches and algorithms, to see which are best at discovering WTFs, and refine challenges; perhaps even make a real discovery!
\item Phase 3: Mount ASKAP early science challenge,  probably find artefacts (which will aid in ASKAP commissioning), test different approaches and algorithms, to see which are best at discovering WTFs in REAL ASKAP data;  increased probability of making a real discovery!
\item Phase 4: By 2017 the EMU survey should be well underway, so we can start adding the EMU data to the WTF machinery. At this stage there is an increasing likelihood of making a real discovery.
\end{itemize}

\section{Conclusion}
WTF is the first stage of our long-term goal of  an extensive international collaboration to apply machine learning and other data science techniques to mine large survey data for new scientific knowledge. 

Our early goals include: (a) building infrastructure and  techniques for mining the unexpected from astronomical data, and
(b) developing experience in using cloud computing as a collaborative research environment.
We also plan to develop, with our collaborators, a set of best practice guidelines for:
\begin{itemize}
\item inviting collaborators to a project like this, including expectation management
\item allocation of resources to collaborators
\item monitoring the usage of and tuning allocations
\item a publishing model for results generated from this work
\end{itemize}

\section*{Acknowledgements}
We thank Amazon Web Services (AWS) for providing a grant to kick-start WTF. We thank all members of the WTF team for contributing to this project.

\bibliography{adassXXVreferences}  

\end{document}